\documentclass[aps,prd,twocolumn,floatfix,nofootinbib,superscriptaddress]{revtex4-1}
\usepackage[colorlinks,linkcolor=blue,anchorcolor=blue,citecolor=blue,urlcolor=blue,breaklinks=true]{hyperref}
\usepackage[libertine]{newtxmath}
\usepackage{graphicx}
\usepackage{slashed}
\usepackage{amsfonts}
\usepackage{amsmath}
\usepackage{blindtext}
\usepackage{microtype}
\usepackage{textcomp}
\usepackage{comment}

\newcommand{\sh}[1]{\slashed{#1}}

\newcommand{\CDOT}{\mbox{$ \cdot $}}
\newcommand{\nslash}{\mbox{$\not \! n$}}

\newcommand{\kslash}{\mbox{$\not \! k$}}

\newcommand{\aslash}{\mbox{$\not \! a$}}

\newcommand{\Eq}[1]{Eq.~\eqref{#1}}

\newcommand{\Fig}[1]{Fig.~{\ref{#1}}}

\newcommand{\Table}[1]{Table~\ref{#1}}

\begin{document}

\title{An imprint of intrinsic quark--gluon correlations: a nonmonotonic feature in $e(x)$.}

\author{Chao Shi}
\email[]{cshi@nuaa.edu.cn}
\affiliation{Department of Nuclear Science and Technology, Nanjing University of Aeronautics and Astronautics, Nanjing 210016, China}

\author{Liming Lu}
\affiliation{Department of Nuclear Science and Technology, Nanjing University of Aeronautics and Astronautics, Nanjing 210016, China}

\author{Ian Clo\"et}
\email[]{icloet@anl.gov}
\affiliation{Physics Division, Argonne National Laboratory, Lemont, IL 60439, USA}

\author{Wenbao Jia}
\affiliation{Department of Nuclear Science and Technology, Nanjing University of Aeronautics and Astronautics, Nanjing 210016, China}

\author{Peter Tandy}
\email[]{tandy@kent.edu}
\affiliation{Center for Nuclear Research, Department of Physics, Kent State University, Kent OH 44242 USA}

\begin{abstract}
We present the first rainbow--ladder Dyson–Schwinger equations study of the pion’s chiral-odd, twist-3 parton distribution $e_{\rm q}(x)$. By deriving a novel Dyson–Schwinger equation for the pion’s quark–quark correlation matrix, we simultaneously extract from it both the unpolarized twist-2 parton distribution function $f_{\rm q}(x)$ and the twist-3 distribution $e_{\rm q}(x)$. Our results show that chiral symmetry strongly suppresses the  twist-2 component of  $e_{\rm q}(x)$, leaving the genuine twist-3 quark–gluon component dominant and  featuring a node.  We therefore argue that the genuine twist-3 term can make a substantial contribution to hadronic $e(x)$, producing a nonmonotonic structure that is a clear imprint of intrinsic quark--gluon correlations. We point out that the existence of a hump-like feature is compatible with, although not uniquely indicated by, recent proton extractions and awaits more precise determination.
\end{abstract}
\maketitle

\noindent\textbf{Introduction:} Parton distribution functions (PDFs) encode the nonperturbative quark and gluon structure of hadrons and are probed in high-energy inclusive and semi-inclusive scattering processes. They comprise a variety of unpolarized and polarized distributions, with or without transverse-momentum dependence, and appear at different orders in the \(1/Q\) expansion, where \(Q\) is the hard scale of the process. Leading contributions scale as \(1/Q^0\) and originate from twist-2 QCD operators, while power-suppressed corrections (\(1/Q, 1/Q^2,\) etc.) arise from operators of higher twist. The unpolarized quark and gluon PDFs, \(f(x)\) and \(g(x)\), are twist-2 and have a clear single-particle probabilistic interpretation. In contrast, higher-twist distributions involve multiparton correlations, offering deeper insight into hadron structure but posing greater theoretical challenges~\cite{Jaffe:1991ra,Efremov:2002qh}. Upcoming measurements, including those at the planned Electron-Ion Collider, are expected to shed new light on these higher-twist effects~\cite{AbdulKhalek:2021gbh,Anderle:2021wcy}. Herein we study the chiral-odd twist-3 unpolarized distribution \(e_\textrm{q}(x)\) of the pion, aiming to illuminate the structure of twist-3 distributions in general.   Possible extension to the nucleon \(e_\textrm{q}(x)\) distribution is considered, for which preliminary data are already available.

\medskip
\noindent\textbf{The twist-3 PDF $e_{\rm q}(x)$:} 
The unpolarized twist-3 PDF $e_{\rm q}(x)$ for quark flavor ${\rm q}$ in hadron $h$, in Minkowski metric, is defined as~\cite{Jaffe:1991kp,Jaffe:1991ra}
\begin{align}\label{eq:exdefALT}
e_{\rm q}(x)=\frac{P\CDOT n}{M_{\rm h}} \int \frac{\textrm{d} \lambda}{4\pi} \textrm{e}^{ix P\cdot n\lambda} \langle P | \bar{\psi}_{\rm q}(0) \rm{I}_4 {\cal W}[0,\lambda n] \psi_{\rm q}(\lambda n) |P \rangle~,
\end{align}
where ${\cal W}[0,\lambda n]$ is the Wilson line gauge link.  This Lorentz invariant form is facilitated by the light cone basis vector $n$, which in the target rest frame is \mbox{$n = (1, -1; \vec{0})$}, and thus \mbox{$a \CDOT n = a^+ = a^0 + a^3$}~\cite{Jaffe:1991kp,Jaffe:1991ra}. 
This twist-3 PDF encodes novel features that are absent in twist-2 PDFs such as the unpolarized quark PDF $f_{\rm q}(x)$.   The latter is given by replacing the unit matrix $\rm{I}_4$ by $\gamma \CDOT n$ and removing the prefactor $ P\CDOT n/M_{\rm h}$.  
To separate the genuine twist-3 and twist-2 components of  \Eq{eq:exdefALT}, an expansion  of the non-local quark field current operator about its local limit (\mbox{$\lambda = 0$}) has been explored~\cite{Belitsky:1997zw,Efremov:2002qh}  through use of the QCD equation of motion for the quark field. This leads to the gauge invariant decomposition~\cite{Belitsky:1997zw,Efremov:2002qh}
\begin{align} \label{eq:exdecomp}
e_{\rm q}(x)=\frac{\langle h| \bar{\psi}_{\rm q}(0)\psi_{\rm q}(0)|h \rangle}{2 M_{\rm h}}\delta(x) + e_{\rm q}^{\textrm{mass}}(x)+e_{\rm q}^{\textrm{tw3}}(x)~,
\end{align}
in which the pure quark term $e_{\rm q}^{\textrm{mass}}(x)$ is twist-2, while $e_{\rm q}^{\textrm{tw3}}(x)$ is generated from quark–gluon correlations and is genuinely twist-3. Some exact features of the three terms have been derived~\cite{Efremov:2002qh} in terms of their  Mellin moments \mbox{$\langle x^m \rangle_e = \int dx x^m e(x) $} which produce
\begin{align} \label{eq:moms0}
m = 0~\Rightarrow~\langle x^0\rangle_{e_{\rm q}} = \frac{\sigma_{\rm q}}{m_{\rm q}}~;~ \langle x^0\rangle_{e_{\rm q}^{\textrm{mass}}} = \langle x^0\rangle_{e_{\rm q}^{\textrm{tw3}}} =0~,
\end{align}
and 
\begin{align} \label{eq:moms1}
m = 1~\Rightarrow~\langle x \rangle_{e_{\rm q}} =  \langle x \rangle_{e_{\rm q}^{\textrm{mass}}} = \frac{m_{\rm q}}{M_{\rm h}}~;~ \langle x \rangle_{e_{\rm q}^{\textrm{tw3}}} =0~,
\end{align}
and 
\begin{align} \label{eq:moms_nge2}
m\ge 2~\Rightarrow~\langle x^m \rangle_{e_{\rm q}^{\textrm{mass}}} = \frac{m_{\rm q}}{M_{\rm h}} \langle x^{m-1} \rangle_{f_{\rm q}}~;~ \langle x^m \rangle_{e_{\rm q}^{\textrm{tw3}}} \neq 0~.
\end{align} 
It is important to note that the $m_{\rm q}$ in the above exact results is the current quark mass. 
We use $\sigma_{\rm q}$ to denote the quark flavor ${\rm q}$ contribution to the hadron sigma term: \mbox{$\sigma_{\rm u} + \sigma_{\rm d}$} gives $\sigma_{\pi {\rm N}}$ for the nucleon, and $\sigma_{\pi}$ for the $\pi^+$. 

Note that although the above is consistent with all moments of $x e_{\rm q}^{\rm mass}(x) = \frac{m_{\rm q}}{M_{\rm h}} f_{\rm q}(x)$, 
the expression $\frac{m_{\rm q}}{M_{\rm h}} f_{\rm q}(x)/x$ does not accurately account for $e_{\rm q}^{\textrm{mass}}(x) $ itself, especially at small $x$, since Eq.~(\ref{eq:moms0}) would not be satisfied.  
The low moment properties for the distribution $x e_{\rm q}(x)$~\cite{Efremov:2002qh} offer a strong hint at  dominant features for light quarks:  the $x e_{\rm q}^{\textrm{mass}}(x)$ term should be a minor contribution due to the relatively small current quark mass; the remaining $x e_{\rm q}^{\textrm{tw3}}(x)$ term integrates to zero, hence it likely changes sign with at least one node, assuming valence-quark dominance over antiquarks. These features are therefore expected to hold for the physical pion, a conclusion supported by our numerical results presented below.

The topic of how to separate the three contributions of Eq.~(\ref{eq:exdecomp}) has attracted extensive discussion in theoretical analyses~\cite{Ma:2020kjz,Bhattacharya:2020jfj,Hatta:2020iin} and nonperturbative models~\cite{Wakamatsu:2003uu,Mukherjee:2009uy,Aslan:2018tff}.
Unfortunately, lattice QCD has not yet constrained \(e_{\rm q}^{\rm tw3}(x)\)~\cite{Bhattacharya:2020jfj}, and existing nucleon extractions of the total  $e_{\rm q}(x)$ from data carry large uncertainties~\cite{Efremov:2002ut,Courtoy:2022kca}.
In this work we calculate  the Mellin moments of the entire $e_{\rm q}(x)$ of the pion using a nonperturbative framework based on the rainbow--ladder truncation of the Dyson--Schwinger equations of QCD.  We use this  DSE-RL approach to obtain the moments of the twist-2 term $x e_{\rm q}^{\rm mass}(x)$ and then identify  the remaining genuine twist-3 term
$x e_{\rm q}^{\textrm{tw3}}(x)$.  The calculation is performed in Euclidean space, where the formulation in terms of integral equations enables an effective resummation of infinite classes of diagrams for two- and three-point functions. The DSE-RL framework offers several key advantages for studying $e(x) $: (i) it retains explicit gluon degrees of freedom,  (ii) it generates dynamical quark mass functions  from current quark masses, and (iii) it preserves chiral symmetry, which could strongly influence $x e_{\rm q}^{\rm mass}(x)$ and thus the deduced $x e_{\rm q}^{\textrm{tw3}}(x)$. 

  \begin{figure}[htbp]
	\centering
	\includegraphics[width=3in]{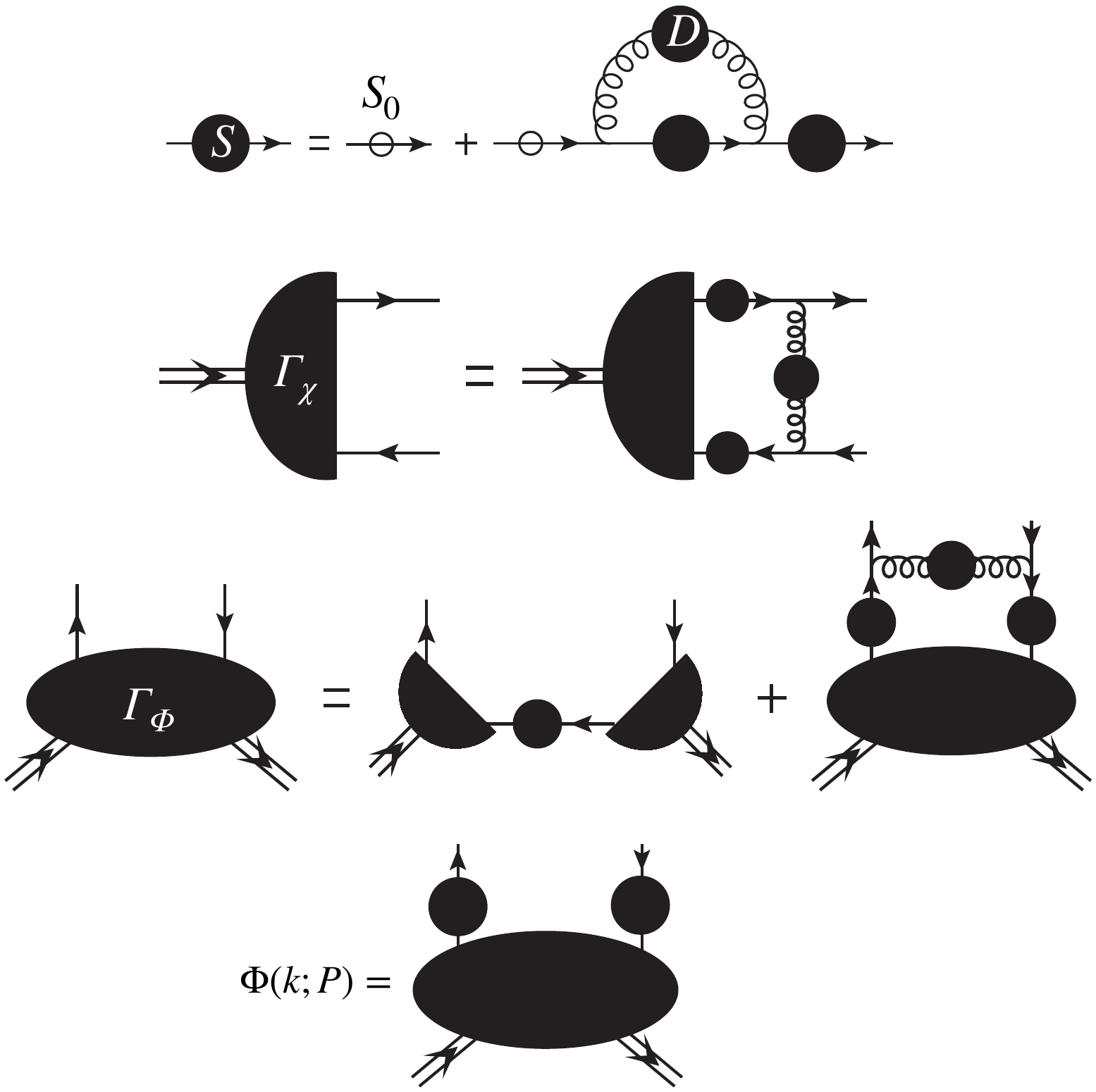}
	\caption{The first three rows are the rainbow--ladder truncated DSEs for dressed quark propagator $S$, the amputated Bethe-Salpeter bound state vertex $\Gamma_{\chi}(k;P)$, and the amputated quark-quark correlation matrix  $\Gamma_\Phi(k;P)$ respectively. The fourth row relates the unamputated quark-quark correlation matrix $\Phi(k;P)$ to $\Gamma_\Phi(k;P)$.}
	\label{fig:DSE}
\end{figure}

\medskip
\noindent\textbf{Pion's quark-quark correlation matrix and rainbow--ladder DSE approach:} 
To facilitate a numerical momentum space DSE-RL treatment of several different types of PDF for a hadron we omit the gauge link and recast Eq.~(\ref{eq:exdefALT}) into the equivalent 2-step form \cite{Jaffe:1983hp,Barone:2001sp} 
\begin{align} \label{eq:phidef}
\Phi_{ij}(k,P)=\int \textrm{d}^4 \xi \ e^{ik_\eta \cdot \xi}\langle P|\bar{\psi}_{j}(0)\psi_{i}(\xi)|P\rangle~,	
\end{align}
followed by 
\begin{align}\label{eq:fdef}
{\rm F}(x)=\frac{1}{2}\int\frac{\textrm{d}^4 k}{(2\pi)^4}\delta(k_\eta\CDOT n - xP\CDOT n)\textrm{Tr}\left[\Phi(k,P)\Gamma_{\rm F} \right]~.
\end{align}
Here $ \Phi_{ij}(k,P)$ is the quark-quark correlator matrix for the hadron and ${\rm F}(x)$ is the PDF associated with  the choice of Dirac matrix $ \Gamma_{\rm F}$. 
The PDFs $f_{\rm q}(x)$ and $e_{\rm q}(x)$ follow from $\Phi_{ij}(k,P)$ 
with use of $\Gamma_{\rm F}=\nslash$ and $\Gamma_{\rm F}=I_4\frac{P\CDOT n}{M_\pi}$ respectively. Note we have chosen $k_\eta \equiv k + \eta P$ as the momentum of the struck quark parton. 

\begin{table*}[ht]
\begin{center}
\begin{tabular}{c@{\hspace{0.8cm}}c@{\hspace{0.8cm}}c@{\hspace{0.8cm}}c@{\hspace{0.8cm}}c@{\hspace{0.8cm}}c@{\hspace{0.8cm}}c@{\hspace{0.8cm}}c@{\hspace{0.8cm}}c@{\hspace{0.8cm}}c@{\hspace{0.8cm}}}
\hline
 $n$  & $0$ & $1$ & $2$ & $3$ & $4$ & $5$ & $6$ & $7$ & $8$ \\\hline
$\langle x^m\rangle_{f_{\rm q}(x)}$ & 1.00 & 0.372 & 0.202 & 0.129 & 0.0910 & 0.0680 & 0.0531 & 0.0428 & 0.0358 \\
$\langle x^m\rangle_{e_{\rm q}(x)}$ &  6.45 & 0.113 & 0.303 & 0.278  & 0.231 & 0.191 & 0.159 & 0.134 & 0.115 \\
$\langle x^m\rangle_{e_{\rm q}^\textrm{mass}(x)}$ & 0  & 0.113 & 0.0419 & 0.0228  & 0.0146 & 0.0103 & 0.00773 & 0.00604 & 0.00488 \\
$\langle x^m\rangle_{e_{\rm q}^{\textrm{tw3}}(x)}$ & 0  & 0 & 0.261 & 0.255  & 0.216 & 0.181 & 0.151 & 0.128 & 0.110 \\\hline
\end{tabular}
\end{center}
\vspace*{-4ex}
\caption{Computed Mellin moments of unpolarized pion PDFs $f_{\rm q}(x)$ for twist-2 and $e_{\rm q}(x)$ for twist-3. Two components of $e_{\rm q}(x)$ are shown and we note that $e_{\rm q}^\textrm{mass}(x)$ is an estimated maximum as discussed in the text.
\label{tab:moms}
}
\end{table*}

The rainbow--ladder truncation preserves the chiral symmetry of QCD and facilitates the associated consequences such as its dynamical breaking to generate constituent quark mass functions.  It also implements  color singlet vector current conservation and the partial conservation of the color singlet axial vector current.
The DSE-RL approach to QCD consists of truncated equations of motion (EOM) for generalized  Green functions 
$\langle\Omega|\hat{O}(\psi,\bar{\psi},A)|\Omega\rangle$~\cite{Maris:1997tm,Maris:2003vk}.  An overview of  the QCD content appropriate to the present task  is illustrated in Fig.~\ref{fig:DSE}.  There row 1  displays the dressed quark propagator \mbox{$S = \langle\Omega|\hat{T}\{\psi(x)\bar{\psi}(y)\}|\Omega\rangle$} as given by the rainbow--ladder integral equation.  Row 2 displays the integral equation for the Bethe-Salpeter (BS) bound state amputated vertex $\langle\Omega|\hat{T}\{\psi(x)\bar{\psi}(y)\}|P\rangle$ for the pion.  The 4-point function representing the amputated pion quark-quark correlation  matrix  $\Gamma_\Phi$ is depicted in row 3.  The unamputated version $\Phi$, defined in Eq.~(\ref{eq:phidef}), is depicted  in row 4.  They are related by $\Phi=S\, \Gamma_\Phi \,S$.   

Detailed momentum space representations within the DSE-RL truncation are as follows.  The dressed quark propagator is obtained from  
\begin{align}
S(k)^{-1} = Z_2 \,(i \kslash &+ Z_4 m_{\rm q}(\mu)) \nonumber \\ 
 \hspace{10mm} &+ Z_2^2\!\! \int^\Lambda_q\!\! g^2 D_{\mu\nu}(q)
\frac{\lambda^a}{2}\gamma_\mu S(k-q) \frac{\lambda^a}{2}\gamma_\nu~, \label{eq:quarkDSE}
\end{align}
which is equivalent to the first row of Fig.~\ref{fig:DSE}, with the open circle $S_0$ denoting the bare quark propagator.   Here $\int^\Lambda_q$ represents 
\mbox{$\int d^4 q/(2 \pi)^4$} with  a smooth Poincar\'e invariant ultraviolet regularization at  mass-scale $\Lambda$.  The $m_{\rm q}(\mu)$ is the current quark mass renormalized at scale $\mu$. The $Z_{2}$ and $Z_4$ are the quark field and mass renormalization constants respectively. The factor $Z_2^2$ preserves multiplicative renormalizability in solutions of the DSE and BSE~\cite{Bloch:2002eq}.  The $D_{\mu\nu}$ is in  Landau gauge and it models a combination of quark-gluon vertex and gluon propagator~\cite{Maris:1997tm,Maris:2003vk}, with details specified later.
The BS bound state vertex is obtained via the integral equation 
\begin{align}
\Gamma_{\chi}(k;P) &=\!-Z_2^2 \!\!\int_q^\Lambda\!\!
\!\!g^2 D_{\mu\nu}(k\!-\!q) \,\gamma^{a}_{\mu} \,S(q_{\eta})\Gamma_{\chi}(q;P) S(q_{\bar{\eta}})\, \gamma^{a}_{\nu}~, \label{eq:mesonBSE}
\end{align}
with \mbox{$ \gamma^{a}_{\mu} \equiv $}  \mbox{$  \frac{\lambda^a}{2}\gamma_\mu$}, and illustrated in the second row of Fig.~\ref{fig:DSE}.  The quark momenta are $q_\eta=q+\eta P$ and $q_{\bar{\eta}}=q-(1-\eta)P$, with $\eta=\frac{1}{2}$ used throughout this work.  

The amputated pion quark-quark correlation  matrix 
$\Gamma_\Phi$ is the 4-point function obtained via the integral equation
\begin{align}
\Gamma_\Phi(k;P) &=\Gamma_\chi (k;P)\, S(k_{\bar{\eta}})\,\bar{\Gamma}_\chi(k;-P) \nonumber\\
&-Z_2^2\!\! \int_q^\Lambda\!\!
g^2 D_{\mu\nu}(k\!-\!q) \gamma^{a}_{\mu}  S(q_\eta)\Gamma_\Phi(q;P) S(q_\eta) \gamma^{a}_{\nu} ~,\label{eq:phiDSE}
\end{align}
and illustrated in the third row of Fig.~\ref{fig:DSE}. The first term on the right hand side is the driving term and the second term incorporates all ladder gluon dressings. It is important to point out that this approach, centered around the collection of all gluon ladders into $\Gamma_\Phi$, is quite different from DSE-RL approaches to PDFs that collect all such ladders to calculate a dressed quark vertex.   This latter dressed quark vertex approach, e.g. Refs.~\cite{Tandy:2023zio,Bednar:2018mtf}, has to be carried out separately for each different bare quark vertex Dirac matrix associated with the PDF of interest.  In contrast, the ladder-summed quark-quark correlation  matrix $\Gamma_\Phi$ depends only on the hadron;  it  can be used in Eq.~(\ref{eq:fdef}) with a variety of bare quark vertices $ \Gamma_{\rm F} $ to yield a variety of PDFs.

\medskip
\noindent\textbf{Numerical implementation:} 
Calculations are facilitated by Euclidian metric\footnote{In Euclidean metric we employ $a_4 = i a^0$ for any space-time vector, including $n$, while  \mbox{$\{\gamma_\alpha, \, \gamma_\beta\} =2 \delta_{\alpha \beta} $}  with $\gamma_4 = \gamma^0$.   Hence \mbox{$\aslash \to -i \aslash $} while \mbox{$a \CDOT b \to - a \CDOT b $}.  }. 
In this work we employ for $ g^2 D_{\mu \nu}$ the infrared term of the Maris-Tandy kernel~\cite{Maris:1999nt}\ as subsequently restructured~\cite{Qin:2011dd}.  Since the main features of parton distributions are dominated by soft infrared dynamics, we neglect the ultraviolet tail to facilitate this initial numerical exploration of $e_{\rm q}(x)$ based on the dressed quark-quark correlator. Thus we use
\begin{align} \label{eq:QC}
g^2 D_{\mu \nu}(k)= \frac{8 \pi^2}{\omega^4} D  \, {\rm e}^{-k^2/\omega^2}\, (\delta_{\mu\nu} - \frac{k_\mu k_\nu}{k^2})~.
\end{align} 
Given this super-renormalizable form, the explicit scale dependence of $m_{\rm q}, Z_2, Z_4$ is unnecessary,  and thus $Z_2=Z_4=1$. Model parameters involved are the current quark mass $m_{u/d}=5.5$ MeV, $\omega=0.5$ GeV and $D\omega=(0.82 \textrm{GeV})^3$, which reproduce the  physical pion mass and decay constant as well as the $\rho$ meson and nucleon properties \cite{Qin:2011dd,Qin:2019hgk,Yao:2024uej}.

The general Dirac structure of the relevant functions is as follows: \mbox{$S(k)^{-1} =i\sh{k}\, A(k^2) + B(k^2)$}, while 
the meson bound state vertex $\Gamma_\chi$ from Eq.~(\ref{eq:mesonBSE}) has the form
\begin{align}  \label{eq:gammachidecomp} 
\Gamma_\chi(k;P)=\gamma^5 \left(\sum_{i=1}^4 T_i \mathcal{F}_{\chi,i}(k^2,k\cdot P,P^2)\right)~,
\end{align}
and the amputated pion quark-quark correlation matrix $\Gamma_\Phi$ from Eq.~(\ref{eq:phiDSE})  has the form
\begin{align}  \label{eq:gammaphidecomp}
\Gamma_\Phi(k;P)=\sum_{i=1}^4 T_i \mathcal{F}_{\Phi,i}(k^2,k\cdot P,P^2)~.
\end{align}
We have used Dirac matrix covariants  $T=\{\textrm{I}_4, \sh{P}, \sh{k}, [\sh{k},\sh{P}]\}$ consistent with parity, time reversal and charge conjugation~\cite{Tangerman:1994eh}.  The $A, B$ functions as well as the four  amplitudes ${\cal F}_{\chi/\Phi}=\{E_{\chi/\Phi}, F_{\chi/\Phi}, G_{\chi/\Phi}, H_{\chi/\Phi}\}$ are Lorentz scalars. They can be obtained numerically via a medium sized computer cluster by combining  Eqs.~(\ref{eq:quarkDSE}-\ref{eq:gammaphidecomp}) to produce results for the correlator $\Phi(k;P)=S(k_\eta)\, \Gamma_\Phi(k;P) \,S(k_\eta)$. 

\medskip
\noindent\textbf{Pion twist-2 and twist-3 PDFs:} 
We  obtain the Mellin moments \mbox{$ \langle x^m \rangle_{\rm F} =\int_{-1}^1 dx\, x^m\, {\rm F}(x) $} via the momentum space approach given by Eq.~(\ref{eq:fdef}) which yields 
\begin{align}
\label{eq:Fmoms}
\langle x^m \rangle_{\rm F} = \frac{1}{2 P\CDOT n} \int\frac{\textrm{d}^4 k}{(2\pi)^4} \, \left(\frac{k_\eta \CDOT n}{P\CDOT n} \right)^m \, \textrm{Tr}\left[\Phi(k,P)\Gamma_{\rm F} \right]~,
\end{align}
with Dirac matrix $\Gamma_{\rm F}$ chosen so that ${\rm F}$ represents $f_{\rm q}$ or $e_{\rm q}$. 
The results are displayed in \Table{tab:moms}. The fact that $\langle x^m \rangle_{e_{\rm q}}$ decreases rapidly  between $m=0$ and $m=1$ is a  signal of significant strength in $e_{\rm q}(x)$ near $x=0$.  The analysis~\cite{Belitsky:1997zw,Efremov:2002qh} which underlies  Eq.~(\ref{eq:exdecomp}), and the moment properties,  begins from the local limit of the bilocal quark field operator in Eq.~(\ref{eq:exdefALT}).  This emphasizes  low $x$ behavior by producing the $\delta(x)$ term which generates the complete  0-th moment $ \sigma_{\pi}/2m_{\rm q}$. 
Our DSE-RL approach yields $\langle x^0 \rangle^{\pi}_{e_{\rm q}}=6.45$ which is equivalent to $\sigma_{\pi}=0.071$ GeV.  This is  essentially $M_\pi/2$, an analytic $\sigma_{\pi}$ result that follows from DSE  rainbow--ladder truncation~\cite{Maris:1999nt,Flambaum:2005kc} combined with the leading order departure of $M_\pi$ from the chiral limit.   

The $\delta(x)$ term in $e_{\rm q}(x)$ does not contribute to hard scattering processes; it is always $xe_{\rm q}(x)$ that shows up in factorized cross sections~\cite{Mulders:1995dh,Hayward:2021psm,Courtoy:2022kca}.   
It is therefore useful to consider $xe_{\rm q}(x)$ as the distribution of interest; all of its moments are equivalent to the $\langle x^m \rangle_{\rm e_{\rm q}}$  for $m \geq 1$.   Our DSE-RL results displayed in \Table{tab:moms} indicate that the distribution $xe_{\rm q}(x)$ has  moments $\langle x^m \rangle^{\pi}_{xe_{\rm q}}$ that do not monotonically decrease with increasing $m$; there is a significant rise from \mbox{$\langle x^0 \rangle^{\pi}_{xe_{\rm q}} = 0.113$}  to   \mbox{$\langle x^1 \rangle^{\pi}_{xe_{\rm q}} = 0.303$}, followed by a slow decrease thereafter. This suggests  a distribution that changes sign  from negative at small $x$ to  positive at large $x$ with at least one node.  We find that if $ xe^{\pi}_{\rm q}(x)$ is parameterized as
\begin{align}
\label{eq:fitxe}
 xe^{\pi}_{\rm q}(x) = N (1-x)^b \left[ x^a - 2 x^c (1-x)  \right],
\end{align}
then an excellent fit to our DSE-RL  moments displayed in \Table{tab:moms} is obtained with deviation less than 0.5\%.  
We also parameterize the twist-2 pion PDF via~\cite{Shi:2024laj}
\begin{align}
\label{eq:fitf}
f^{\pi}_{\rm q}(x)=  (1-x)^\beta \sum_{i=0,1,2}^{\ } c_i x^i,
\end{align}
and the DSE-RL moments given in \Table{tab:moms} are fit with deviation less than 1\%.   The resulting parameters are listed in \Table{tab:para}.  

We note from Eqs.~(\ref{eq:exdecomp},\ref{eq:moms1}) that an exact QCD result for parton momentum is \mbox{$\langle x^0\rangle_{xe_{\rm q}} \equiv$}  \mbox{$\langle x \rangle_{e_{\rm q}} = $} \mbox{$\langle x \rangle_{e_{\rm q}^{\textrm{mass}}}$} \mbox{$  = m_{\rm q}/M_\pi$}.  With our current mass value $m_{\rm q}=5.5$ MeV, this would give $0.039$, but we obtain $0.113$.  
The difference arises because the DSE-RL truncation approach, based on ladder sums of quark-quark interactions, can not exactly preserve the transformation of the quark-quark correlation under the QCD EOM for the quark field operator that underlies the above. However, our $\langle x^0\rangle_{xe_{\rm q}}$ is still much smaller than the next moment \mbox{$\langle x \rangle^{\pi}_{xe_{\rm q}} = 0.303$} and that dominates the consequent $x$-dependence. 
Moreover, we have verified that $\langle x\rangle_{e_{\rm q}} \to 0$ as
$m_{\rm q} \to 0$, confirming the strong chiral-symmetry constraint on the pion’s $e(x)$,
consistent with the Gell-Mann--Oakes--Renner relation $m_{\rm q} \propto M_\pi^2$ near the
chiral limit.

\begin{table}[h!]

\begin{center}
\begin{tabular*}
{\hsize}
{
c@{\extracolsep{0ptplus1fil}}|
c@{\extracolsep{0ptplus1fil}}
c@{\extracolsep{0ptplus1fil}}
c@{\extracolsep{0ptplus1fil}}
c@{\extracolsep{0ptplus1fil}}}\hline
 $f_{\rm q}^\pi(x)$  & $\beta$   & $c_0$   & $c_1$    & $c_2$   \\\hline
                              &  0.704     & 1.737    & -0.3587  & 0.4464  \\\hline
  $xe_{\rm q}^\pi(x)$\hspace*{1ex}   &  $N$    &   $a$       &   $b$     & $c$   \\\hline
                                                    &  8.493  & 0.4496   & 0.7528  & 0.5512   \\\hline
\end{tabular*}
\end{center}
\vspace*{-4ex}
\caption{Fitting parameters for the pion PDFs in Eqs.~(\ref{eq:fitxe},\ref{eq:fitf}). 
\label{tab:para}
}
\end{table}

\begin{figure}[htbp]
\centering
\includegraphics[width=3.5in]{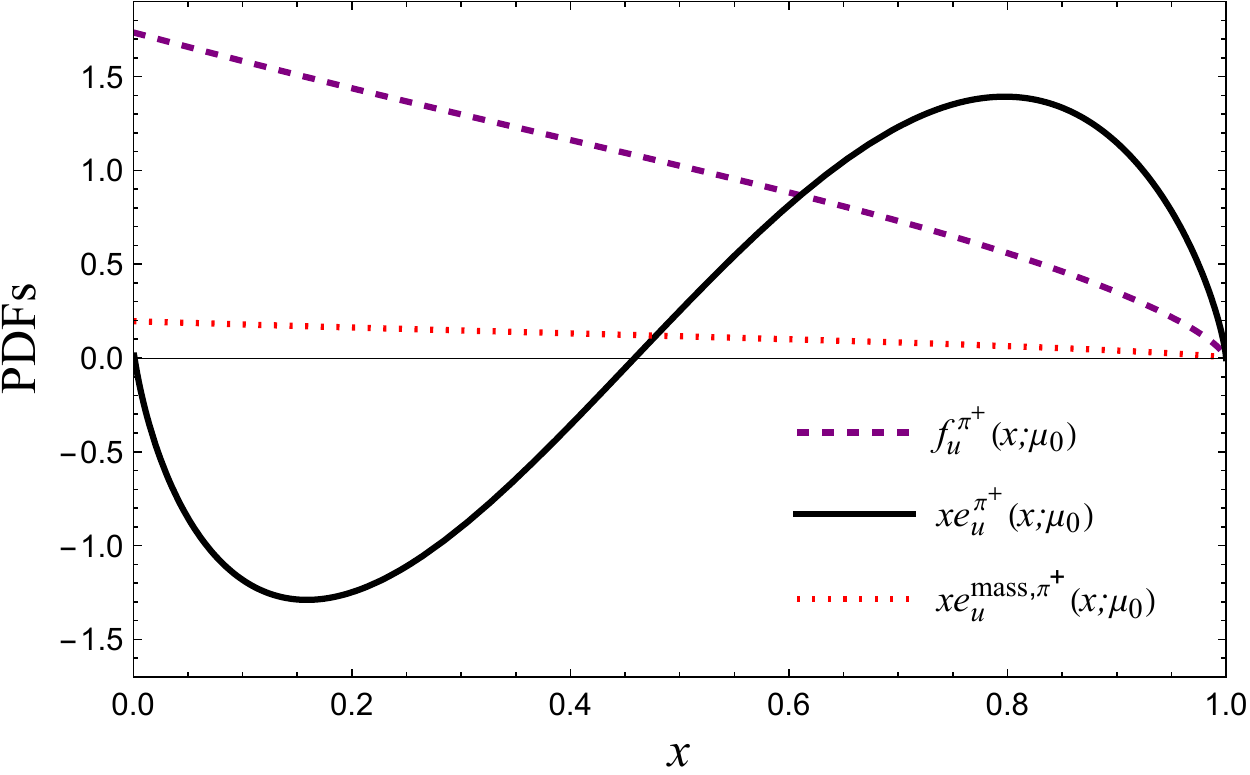}
\hspace*{0.2cm}\includegraphics[width=3.43in]{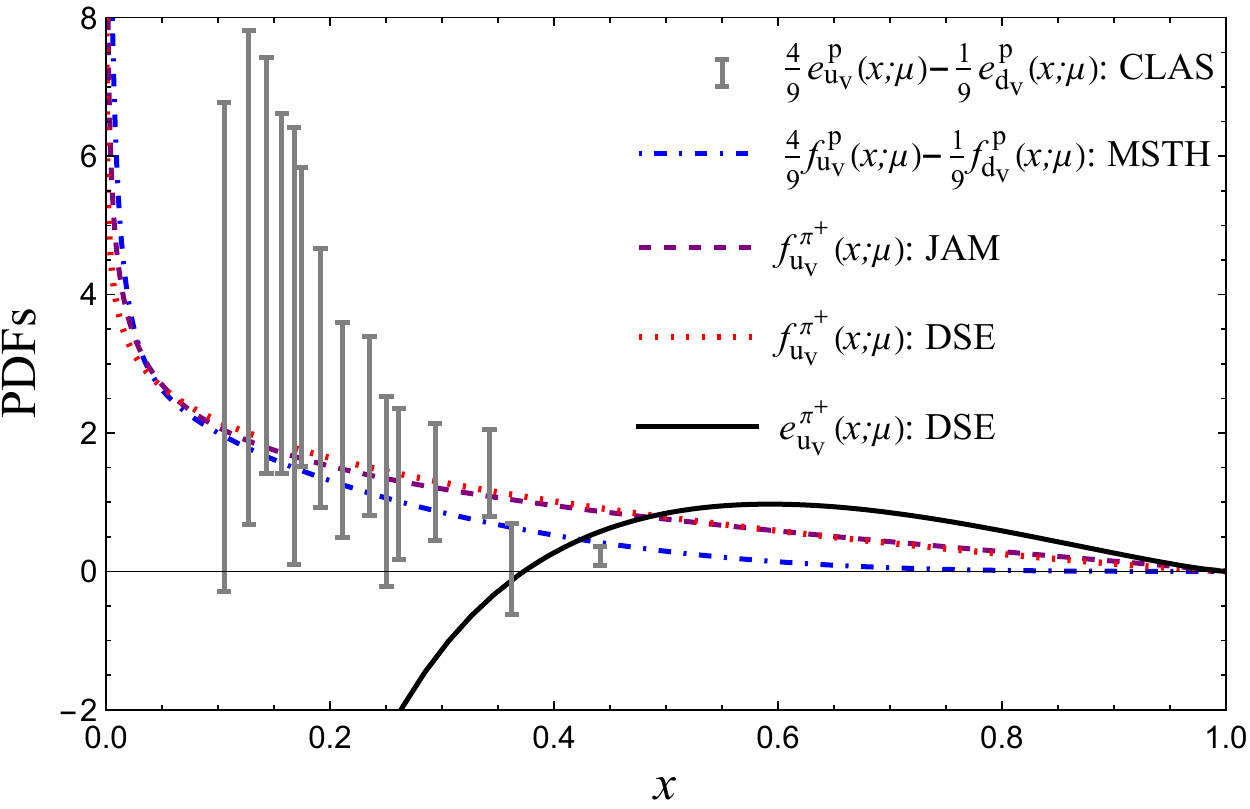}
\caption{ \emph{Upper Panel:} The pion PDFs at model scale $\mu_0=0.63$ GeV.  \emph{Lower Panel:} The pion PDFs $f_{\rm q}(x)$ and 
$e_{\rm q}(x)$ at $\mu =1.3$ GeV with $f_{\rm q}(x)$  compared to the JAM analysis at NLO~\cite{Barry:2018ort}.  Also shown for comparison are isospin 
averaged proton PDF data: the  $e_{\rm q}(x)$ CLAS data points at approximately $1$ GeV~\cite{Courtoy:2022kca}, and the MSHT fit of the proton $f_{\rm q}(x)$~\cite{Bailey:2020ooq}.}
\label{fig:PDFs}
\end{figure}

The considered DSE-RL diagrams in Fig.~\ref{fig:DSE} only constitute a subset of all possible QCD diagrams, e.g., quark–antiquark pair creation and annihilation from the vacuum are ignored. This associates a model scale $\mu_0$ to our approach, at which intrinsic sea quark PDFs are absent, i.e., \mbox{$f_{\bar{\rm q}}(x;\mu_0)=e_{\bar{\rm q}}(x;\mu_0)=0$}. Given $f_{\rm q}(-x) = - f_{\bar{\rm q}}(x)$ and $e_{\rm q}(-x) = e_{\bar{\rm q}}(x)$ by definition~\cite{Jaffe:1991ra,Efremov:2002qh}, the obtained PDFs $f_{\rm q}(x;\mu_0)$ and $xe_{\rm q}(x;\mu_0)$ are thus only nonzero in the domain $x\in [0,1]$.  The results from fitting the moments are  shown in the upper panel of Fig.~\ref{fig:PDFs}. The model scale  $\mu_0 \approx 0.63$ GeV is determined after QCD evolution as described below. Noticeably, the $xe_{\rm q}(x)$ is comparable with $f_{\rm q}(x)$ in magnitude. Regarding the decomposition in Eq.~(\ref{eq:exdecomp}), we estimate for the pion  $x e_{\rm q}^{\rm mass}(x)\equiv 0.113 f_{\rm q}(x)$ which saturates the present DSE-RL result $\langle x\rangle_{e^{ \pi}_{\rm q}}=0.113$.   
As a result Fig.~\ref{fig:PDFs} shows that the resulting genuine twist-3 contribution $x e_{\rm q}^{\textrm{tw3}}(x) =x e_{\rm q}(x) - x e_{\rm q}^{\textrm{mass}}(x)$ is strongly dominant.  Moreover, due to the sum rule constraint \mbox{$\langle x^0 \rangle_{x e_{\rm q}^{\textrm{tw3}}} = 0$}, the $x$ shape of $xe_{\rm q}^{\textrm{tw3}}(x)$ is oscillatory, analogously for the total $xe_{\rm q}(x)$. The Mellin moments of the estimated pion $e_{\rm q}^{\textrm{mass}}(x)$ and the deduced $e_{\rm q}^{\textrm{tw3}}(x)$ are displayed in  \Table{tab:moms}.

The QCD evolution equations of $e_{\rm q}(x)$ were developed in \cite{Koike:1996bs,Balitsky:1996uh,Belitsky:1997zw}. At large $N_c$ and in the chiral limit, the moments of the valence $e_{\rm q}(x)$ obey the simple DGLAP equation similar to $f_{\rm q}(x)$
\begin{align}
\langle x^m\rangle_{f_{\rm q}}(\mu)&=L^{\gamma_n^f/\beta_0}\langle x^m\rangle_{f_{\rm q}}(\mu_0), \ \ \  n\ge 0 \label{eq:evo1}\\
\langle x^m\rangle_{e_{\rm q}}(\mu)&=L^{\gamma_n^e/\beta_0}\langle x^m\rangle_{e_{\rm q}}(\mu_0), \ \ \  n\ge 2 \label{eq:evo2}
\end{align}
with $L\equiv\alpha_s(\mu)/\alpha_s(\mu_0)$, $\gamma_n^e=2N_c\left(S_n-\frac{1}{2(n+1)}-\frac{1}{4}\right)$, $\gamma_n^f=\frac{4}{3}\left(4 S_{n+1}-3-\frac{2}{(n+1)(n+2)}\right)$, $S_n\equiv\sum_{i=1}^{n}\frac{1}{i}$ and $\beta_0=11 – \frac{2}{3}N_f$. 
\Eq{eq:evo2} was found to be a good approximation to the exact one \cite{Koike:1996bs}.  
Here at leading order we take $N_f=3$ and $\Lambda_{\textrm{QCD}}=0.35$ GeV and evolve the PDFs to the scale of $\mu=1.3$ GeV. The initial scale $\mu_0$ is determined by matching to the JAM global fit result $\langle x\rangle_{f_{\textrm{q}_\textrm{v}}}=0.268$ \cite{Barry:2018ort}. Note the valence combination is defined as $f_{\textrm{q}_\textrm{v}} \equiv f_{\textrm{q}} - f_{\bar{\textrm{q}}}$, and analogously for $e_{\textrm{q}_\textrm{v}}$. 
 All other moments of $f_{\rm q}(x;\mu)$ and $e_{\rm q}(x;\mu)$ are thus predictions. 

The evolved valence PDFs of the pion are shown in the lower panel of \Fig{fig:PDFs}. 
Our $f^{\pi^+}_{\rm u_{\rm v}}(x;\mu)$ compares well with the JAM analysis at NLO~\cite{Barry:2018ort}, and applying NLO evolution produces only minor modifications. 
Meanwhile, $e^{\pi^+}_{\rm u_{\rm v}}(x;\mu)$ exhibits two salient features: (i) its magnitude remains comparable to that of the twist-2 PDF $f^{\pi^+}_{\rm u_{\rm v}}(x;\mu)$; (ii) the zero-crossing node evident at model scale persists, albeit shifted toward smaller $x$.

A rainbow--ladder DSE computation of the nucleon’s $e(x)$ can be carried out analogously by extending the method developed in this work, albeit at a significantly higher computational cost. In this case, however, the dominance of the genuine twist-3 contribution $e_{\rm q}^{\rm tw3}(x)$ in the proton may be less pronounced at hadronic scales, since effective constituent-like quark degrees of freedom become relevant in $x e_{\rm q}^{\rm mass}(x)=\frac{m_{\rm q}}{M_{\rm h}}f(x)$, as chiral symmetry is less constraining than in the pion \footnote{We emphasize that this discussion is restricted to a model or effective-theory calculation at a hadronic scale. At higher experimental scales, proton's $e_{\rm q}^{\rm mass}(x)$ is strongly suppressed by the factor $\frac{m_{\rm q}}{M_{\rm proton}}$, so that only $e_{\rm q}^{\rm tw3}(x)$ could potentially play a dominant role.}. In various model studies, the effective quark mass is found to lie in the range $100$--$400$ MeV~\cite{Efremov:2002qh,Schweitzer:2003uy,Ohnishi:2003mf,Pasquini:2018oyz,Zhu:2024awq}, reflecting the enhancement induced by infrared dynamics at hadronic scales. Nevertheless, it remains possible that $e_{\rm q}^{\rm tw3}(x)$ still contributes substantially. If so, its nodal structure would leave a characteristic imprint---a nonmonotonic hump in $e(x)$ at experimental scales. Such a feature would be absent if $e(x)$ were dominated by the mass term $e_{\rm q}^{\rm mass}(x)$ and its evolution pattern. Figure~\ref{fig:PDFs} shows the isospin-averaged proton $e_{\rm q}(x)$ extracted from CLAS~\cite{Courtoy:2022kca} together with the central MSHT20nlo fit of $f_{\rm q}(x)$~\cite{Bailey:2020ooq}. A hump-like feature is compatible with, although not uniquely indicated by, the data, given the sizable experimental uncertainties.

\medskip
\noindent\textbf{Summary and Outlook:} 
We present the first rainbow--ladder Dyson--Schwinger study of the pion’s twist-3 PDF $e_{\rm q}(x)$. 
The  Mellin moments are computed first and the $x$ distribution is derived from them.  The results strongly support the QCD equation-of-motion decomposition of \Eq{eq:exdecomp}, exhibiting the $\delta(x)$ singularity and a chiral-symmetry suppression of the twist-2 term $e_{\rm q}^{\rm mass}(x)$. 
These findings provide a new picture for the pion  $e_{\rm q}(x)$: the twist-3 quark--gluon piece, rather than the commonly assumed twist-2 term, dominates the distribution.  
A characteristic feature of the resulting $e(x)$ is a nodal structure and the associated hump.
The presently available data, limited by large uncertainties, can neither pin down nor exclude the existence of a hump. Higher-precision measurements are required to clarify this point, and extending measurements to lower $x$ at an Electron–Ion Collider may determine whether the hump evolves into a zero crossing at lower $x$ \cite{LHeC:2020van,AbdulKhalek:2021gbh,Anderle:2021wcy}.

\noindent\textbf{Acknowledgement:} 
This work was inspired in part by discussions with the late Professor Fan Wang.

\bibliography{piex}

@article{Tandy:2023zio,
    author = "Tandy, Peter C.",
    title = "{Parton decomposition of nucleon spin and momentum: Gluons from dressed quarks}",
    eprint = "2302.07473",
    archivePrefix = "arXiv",
    primaryClass = "hep-ph",
    doi = "10.1016/j.physletb.2023.137972",
    journal = "Phys. Lett. B",
    volume = "842",
    pages = "137972",
    year = "2023"
}

@article{Jaffe:1983hp,
    author = "Jaffe, R. L.",
    title = "{Parton Distribution Functions for Twist Four}",
    reportNumber = "MIT-CTP-1085",
    doi = "10.1016/0550-3213(83)90361-9",
    journal = "Nucl. Phys. B",
    volume = "229",
    pages = "205--230",
    year = "1983"
}

@article{Yao:2024uej,
    author = "Yao, Zhao-Qian and Binosi, Daniele and Cu, Zhu-Fang and Roberts, Craig D.",
    title = "{Nucleon charge and magnetisation distributions: flavour separation and zeroes}",
    eprint = "2403.08088",
    archivePrefix = "arXiv",
    primaryClass = "hep-ph",
    reportNumber = "NJU-INP 085/24",
    month = "3",
    year = "2024"
}

@article{Zhu:2024awq,
    author = "Zhu, Zhimin and Xu, Siqi and Wu, Jiatong and Yu, Hongyao and Hu, Zhi and Lan, Jiangshan and Mondal, Chandan and Zhao, Xingbo and Vary, James P.",
    collaboration = "BLFQ",
    title = "{Transverse structure of the proton beyond leading twist: A light-front Hamiltonian approach}",
    eprint = "2404.13720",
    archivePrefix = "arXiv",
    primaryClass = "hep-ph",
    doi = "10.1016/j.physletb.2024.138829",
    journal = "Phys. Lett. B",
    volume = "855",
    pages = "138829",
    year = "2024"
}

@article{Bailey:2020ooq,
    author = "Bailey, S. and Cridge, T. and Harland-Lang, L. A. and Martin, A. D. and Thorne, R. S.",
    title = "{Parton distributions from LHC, HERA, Tevatron and fixed target data: MSHT20 PDFs}",
    eprint = "2012.04684",
    archivePrefix = "arXiv",
    primaryClass = "hep-ph",
    reportNumber = "IPPP/20/58",
    doi = "10.1140/epjc/s10052-021-09057-0",
    journal = "Eur. Phys. J. C",
    volume = "81",
    number = "4",
    pages = "341",
    year = "2021"
}

@article{Tangerman:1994eh,
    author = "Tangerman, R. D. and Mulders, P. J.",
    title = "{Intrinsic transverse momentum and the polarized Drell-Yan process}",
    eprint = "hep-ph/9403227",
    archivePrefix = "arXiv",
    reportNumber = "NIKHEF-94-P1",
    doi = "10.1103/PhysRevD.51.3357",
    journal = "Phys. Rev. D",
    volume = "51",
    pages = "3357--3372",
    year = "1995"
}

@Article{Maris:1999nt,
     author    = "Maris, Pieter and Tandy, Peter C.",
     title     = "Bethe-Salpeter study of vector meson masses and decay
                  constants",
     journal   = "Phys. Rev.",
     volume    = "C60",
     year      = "1999",
     pages     = "055214",
     eprint    = "nucl-th/9905056",
     SLACcitation  = "%%CITATION = NUCL-TH 9905056;%%"
}

@article{Qin:2011dd,
      author         = "Qin, Si-xue and Chang, Lei and Liu, Yu-xin and Roberts,
                        Craig D. and Wilson, David J.",
      title          = "{Interaction model for the gap equation}",
      journal        = "Phys.Rev.",
      volume         = "C84",
      pages          = "042202",
      doi            = "10.1103/PhysRevC.84.042202",
      year           = "2011",
      eprint         = "1108.0603",
      archivePrefix  = "arXiv",
      primaryClass   = "nucl-th",
      SLACcitation   = "%%CITATION = ARXIV:1108.0603;%%",
}

@article{Hatta:2020iin,
    author = "Hatta, Yoshitaka and Zhao, Yong",
    title = "{Parton distribution function for the gluon condensate}",
    eprint = "2006.02798",
    archivePrefix = "arXiv",
    primaryClass = "hep-ph",
    doi = "10.1103/PhysRevD.102.034004",
    journal = "Phys. Rev. D",
    volume = "102",
    number = "3",
    pages = "034004",
    year = "2020"
}

@article{Koike:1996bs,
    author = "Koike, Yuji and Nishiyama, N.",
    title = "{Q**2 evolution of chiral odd twist - three distribution e (x, Q**2)}",
    eprint = "hep-ph/9609207",
    archivePrefix = "arXiv",
    doi = "10.1103/PhysRevD.55.3068",
    journal = "Phys. Rev. D",
    volume = "55",
    pages = "3068--3076",
    year = "1997"
}

@article{Belitsky:1997zw,
    author = "Belitsky, Andrei V. and Mueller, Dieter",
    title = "{Scale dependence of the chiral odd twist - three distributions h-L(x) and e(x)}",
    eprint = "hep-ph/9702354",
    archivePrefix = "arXiv",
    reportNumber = "CERN-TH-97-25, UL-NTZ-97-6",
    doi = "10.1016/S0550-3213(97)00432-X",
    journal = "Nucl. Phys. B",
    volume = "503",
    pages = "279--308",
    year = "1997"
}

@article{Balitsky:1996uh,
    author = "Balitsky, I. I. and Braun, Vladimir M. and Koike, Y. and Tanaka, K.",
    title = "{Q**2 evolution of chiral odd twist - three distributions h(L) (x, Q**2) and e (x, Q**2) in the large N(c) limit}",
    eprint = "hep-ph/9605439",
    archivePrefix = "arXiv",
    reportNumber = "NORDITA-96-32-P, JUPD-9612, MIT-CTP-2537",
    doi = "10.1103/PhysRevLett.77.3078",
    journal = "Phys. Rev. Lett.",
    volume = "77",
    pages = "3078--3081",
    year = "1996"
}

@article{Jaffe:1991kp,
    author = "Jaffe, R. L. and Ji, Xiang-Dong",
    title = "{Chiral odd parton distributions and polarized Drell-Yan}",
    reportNumber = "MIT-CTP-1952",
    doi = "10.1103/PhysRevLett.67.552",
    journal = "Phys. Rev. Lett.",
    volume = "67",
    pages = "552--555",
    year = "1991"
}

@article{Barry:2018ort,
    author = "Barry, P. C. and Sato, N. and Melnitchouk, W. and Ji, Chueng-Ryong",
    title = "{First Monte Carlo Global QCD Analysis of Pion Parton Distributions}",
    eprint = "1804.01965",
    archivePrefix = "arXiv",
    primaryClass = "hep-ph",
    reportNumber = "JLAB-THY-18-2678",
    doi = "10.1103/PhysRevLett.121.152001",
    journal = "Phys. Rev. Lett.",
    volume = "121",
    number = "15",
    pages = "152001",
    year = "2018"
}

@article{Flambaum:2005kc,
    author = "Flambaum, V. V. and Holl, A. and Jaikumar, P. and Roberts, C. D. and Wright, S. V.",
    title = "{Sigma terms of light-quark hadrons}",
    eprint = "nucl-th/0510075",
    archivePrefix = "arXiv",
    reportNumber = "ANL-PHY-11391-TH-2005, MPG-VT-UR-263-05",
    doi = "10.1007/s00601-005-0123-1",
    journal = "Few Body Syst.",
    volume = "38",
    pages = "31--51",
    year = "2006"
}

@article{Shi:2024laj,
    author = "Shi, Chao and Liu, Pengfei and Du, Yi-Lun and Jia, Wenbao",
    title = "{Heavy flavor-asymmetric pseudoscalar mesons on the light front}",
    eprint = "2409.05098",
    archivePrefix = "arXiv",
    primaryClass = "hep-ph",
    doi = "10.1103/PhysRevD.110.094010",
    journal = "Phys. Rev. D",
    volume = "110",
    number = "9",
    pages = "094010",
    year = "2024"
}

@article{Jaffe:1991ra,
    author = "Jaffe, R. L. and Ji, Xiang-Dong",
    title = "{Chiral odd parton distributions and Drell-Yan processes}",
    reportNumber = "MIT-CTP-2005",
    doi = "10.1016/0550-3213(92)90110-W",
    journal = "Nucl. Phys. B",
    volume = "375",
    pages = "527--560",
    year = "1992"
}

@article{Ohnishi:2003mf,
    author = "Ohnishi, Y. and Wakamatsu, M.",
    title = "{pi N sigma term and chiral odd twist three distribution function e(x) of the nucleon in the chiral quark soliton model}",
    eprint = "hep-ph/0312044",
    archivePrefix = "arXiv",
    reportNumber = "OU-HEP-459",
    doi = "10.1103/PhysRevD.69.114002",
    journal = "Phys. Rev. D",
    volume = "69",
    pages = "114002",
    year = "2004"
}

@article{Schweitzer:2003uy,
    author = "Schweitzer, P.",
    title = "{The Chirally odd twist three distribution function e**alpha(x) in the chiral quark soliton model}",
    eprint = "hep-ph/0303011",
    archivePrefix = "arXiv",
    doi = "10.1103/PhysRevD.67.114010",
    journal = "Phys. Rev. D",
    volume = "67",
    pages = "114010",
    year = "2003"
}

@article{Courtoy:2022kca,
    author = "Courtoy, Aurore and Miramontes, Angel S. and Avakian, Harut and Mirazita, Marco and Pisano, Silvia",
    title = "{Extraction of the higher-twist parton distribution e(x) from CLAS data}",
    eprint = "2203.14975",
    archivePrefix = "arXiv",
    primaryClass = "hep-ph",
    doi = "10.1103/PhysRevD.106.014027",
    journal = "Phys. Rev. D",
    volume = "106",
    number = "1",
    pages = "014027",
    year = "2022"
}

@article{Bednar:2018mtf,
    author = {Bednar, Kyle D. and Clo\"et, Ian C. and Tandy, Peter C.},
    title = "{Distinguishing Quarks and Gluons in Pion and Kaon Parton Distribution Functions}",
    eprint = "1811.12310",
    archivePrefix = "arXiv",
    primaryClass = "nucl-th",
    doi = "10.1103/PhysRevLett.124.042002",
    journal = "Phys. Rev. Lett.",
    volume = "124",
    number = "4",
    pages = "042002",
    year = "2020"
}

@article{Mulders:1995dh,
    author = "Mulders, P. J. and Tangerman, R. D.",
    title = "{The Complete tree level result up to order 1/Q for polarized deep inelastic leptoproduction}",
    eprint = "hep-ph/9510301",
    archivePrefix = "arXiv",
    reportNumber = "NIKHEF-95-053",
    doi = "10.1016/0550-3213(95)00632-X",
    journal = "Nucl. Phys. B",
    volume = "461",
    pages = "197--237",
    year = "1996",
    note = "[Erratum: Nucl.Phys.B 484, 538--540 (1997)]"
}

@article{Bloch:2002eq,
    author = "Bloch, Jacques C. R.",
    title = "{Multiplicative renormalizability and quark propagator}",
    eprint = "hep-ph/0202073",
    archivePrefix = "arXiv",
    reportNumber = "UNITU-THEP-02-02",
    doi = "10.1103/PhysRevD.66.034032",
    journal = "Phys. Rev. D",
    volume = "66",
    pages = "034032",
    year = "2002"
}

@article{Hayward:2021psm,
    author = "Hayward, T. B. and others",
    title = "{Observation of Beam Spin Asymmetries in the Process $ep\rightarrow{e}^{'}{\pi}^{+}{\pi}^{-}X$ with CLAS12}",
    eprint = "2101.04842",
    archivePrefix = "arXiv",
    primaryClass = "hep-ex",
    reportNumber = "JLAB-PHY-21-3307",
    doi = "10.1103/PhysRevLett.126.152501",
    journal = "Phys. Rev. Lett.",
    volume = "126",
    pages = "152501",
    year = "2021"
}

@article{Maris:1997tm,
    author = "Maris, Pieter and Roberts, Craig D.",
    title = "{Pi- and K meson Bethe-Salpeter amplitudes}",
    eprint = "nucl-th/9708029",
    archivePrefix = "arXiv",
    reportNumber = "ANL-PHY-8788-TH-97",
    doi = "10.1103/PhysRevC.56.3369",
    journal = "Phys. Rev. C",
    volume = "56",
    pages = "3369--3383",
    year = "1997"
}

@article{LHeC:2020van,
    author = "Agostini, P. and others",
    collaboration = "LHeC, FCC-he Study Group",
    title = "{The Large Hadron\textendash{}Electron Collider at the HL-LHC}",
    eprint = "2007.14491",
    archivePrefix = "arXiv",
    primaryClass = "hep-ex",
    reportNumber = "CERN-ACC-Note-2020-0002, JLAB-ACP-20-3180",
    doi = "10.1088/1361-6471/abf3ba",
    journal = "J. Phys. G",
    volume = "48",
    number = "11",
    pages = "110501",
    year = "2021"
}

@article{Anderle:2021wcy,
    author = "Anderle, Daniele P. and others",
    title = "{Electron-ion collider in China}",
    eprint = "2102.09222",
    archivePrefix = "arXiv",
    primaryClass = "nucl-ex",
    reportNumber = "Frontiers of Physics, Volume 16 Issue (6):64701, 2021",
    doi = "10.1007/s11467-021-1062-0",
    journal = "Front. Phys. (Beijing)",
    volume = "16",
    number = "6",
    pages = "64701",
    year = "2021"
}

@article{Bhattacharya:2020jfj,
    author = "Bhattacharya, Shohini and Cichy, Krzysztof and Constantinou, Martha and Metz, Andreas and Scapellato, Aurora and Steffens, Fernanda",
    title = "{The role of zero-mode contributions in the matching for the twist-3 PDFs $e(x)$ and $h_{L}(x)$}",
    eprint = "2006.12347",
    archivePrefix = "arXiv",
    primaryClass = "hep-ph",
    doi = "10.1103/PhysRevD.102.114025",
    journal = "Phys. Rev. D",
    volume = "102",
    pages = "114025",
    year = "2020"
}

@article{Efremov:2002qh,
    author = "Efremov, A. V. and Schweitzer, P.",
    title = "{The Chirally odd twist 3 distribution e(a)(x)}",
    eprint = "hep-ph/0212044",
    archivePrefix = "arXiv",
    doi = "10.1088/1126-6708/2003/08/006",
    journal = "JHEP",
    volume = "08",
    pages = "006",
    year = "2003"
}

@article{Mukherjee:2009uy,
    author = "Mukherjee, Asmita",
    title = "{Twist Three Distribution e(x): Sum Rules and Equation of Motion Relations}",
    eprint = "0912.1446",
    archivePrefix = "arXiv",
    primaryClass = "hep-ph",
    doi = "10.1016/j.physletb.2010.03.023",
    journal = "Phys. Lett. B",
    volume = "687",
    pages = "180--183",
    year = "2010"
}

@article{Wakamatsu:2003uu,
    author = "Wakamatsu, M. and Ohnishi, Y.",
    title = "{The Nonperturbative origin of delta function singularity in the chirally odd twist three distribution function e(x)}",
    eprint = "hep-ph/0303007",
    archivePrefix = "arXiv",
    reportNumber = "OU-HEP-433",
    doi = "10.1103/PhysRevD.67.114011",
    journal = "Phys. Rev. D",
    volume = "67",
    pages = "114011",
    year = "2003"
}

@article{Maris:2003vk,
    author = "Maris, Pieter and Roberts, Craig D.",
    title = "{Dyson-Schwinger equations: A Tool for hadron physics}",
    eprint = "nucl-th/0301049",
    archivePrefix = "arXiv",
    reportNumber = "ANL-PHY-10465-TH-2002",
    doi = "10.1142/S0218301303001326",
    journal = "Int. J. Mod. Phys. E",
    volume = "12",
    pages = "297--365",
    year = "2003"
}

@article{Efremov:2002ut,
    author = "Efremov, A. V. and Goeke, K. and Schweitzer, P.",
    title = "{Azimuthal asymmetries at CLAS: extraction of e**a(x) and prediction of A(UL)}",
    eprint = "hep-ph/0208124",
    archivePrefix = "arXiv",
    doi = "10.1103/PhysRevD.67.114014",
    journal = "Phys. Rev. D",
    volume = "67",
    pages = "114014",
    year = "2003"
}

@article{Qin:2019hgk,
    author = "Qin, Si-xue and Roberts, Craig D and Schmidt, Sebastian M",
    title = "{Spectrum of light- and heavy-baryons}",
    eprint = "1902.00026",
    archivePrefix = "arXiv",
    primaryClass = "nucl-th",
    doi = "10.1007/s00601-019-1488-x",
    journal = "Few Body Syst.",
    volume = "60",
    number = "2",
    pages = "26",
    year = "2019"
}

@article{Pasquini:2018oyz,
    author = "Pasquini, Barbara and Rodini, Simone",
    title = "{The twist-three distribution $e^q(x,k_\perp)$ in a light-front model}",
    eprint = "1806.10932",
    archivePrefix = "arXiv",
    primaryClass = "hep-ph",
    doi = "10.1016/j.physletb.2018.11.033",
    journal = "Phys. Lett. B",
    volume = "788",
    pages = "414--424",
    year = "2019"
}

@article{Ma:2020kjz,
    author = "Ma, J. P. and Zhang, G. P.",
    title = "{On the singular behavior of the chirality-odd twist-3 parton distribution $e(x)$}",
    eprint = "2003.13920",
    archivePrefix = "arXiv",
    primaryClass = "hep-ph",
    doi = "10.1016/j.physletb.2020.135947",
    journal = "Phys. Lett. B",
    volume = "811",
    pages = "135947",
    year = "2020"
}

@article{Aslan:2018tff,
    author = "Aslan, Fatma and Burkardt, Matthias",
    title = "{Singularities in Twist-3 Quark Distributions}",
    eprint = "1811.00938",
    archivePrefix = "arXiv",
    primaryClass = "nucl-th",
    doi = "10.1103/PhysRevD.101.016010",
    journal = "Phys. Rev. D",
    volume = "101",
    number = "1",
    pages = "016010",
    year = "2020"
}

@article{Barone:2001sp,
    author = "Barone, Vincenzo and Drago, Alessandro and Ratcliffe, Philip G.",
    title = "{Transverse polarisation of quarks in hadrons}",
    eprint = "hep-ph/0104283",
    archivePrefix = "arXiv",
    doi = "10.1016/S0370-1573(01)00051-5",
    journal = "Phys. Rept.",
    volume = "359",
    pages = "1--168",
    year = "2002"
}

@article{AbdulKhalek:2021gbh,
    author = "Abdul Khalek, R. and others",
    title = "{Science Requirements and Detector Concepts for the Electron-Ion Collider}: {EIC Yellow Report}",
    eprint = "2103.05419",
    archivePrefix = "arXiv",
    primaryClass = "physics.ins-det",
    reportNumber = "BNL-220990-2021-FORE, JLAB-PHY-21-3198, LA-UR-21-20953",
    doi = "10.1016/j.nuclphysa.2022.122447",
    journal = "Nucl. Phys. A",
    volume = "1026",
    pages = "122447",
    year = "2022"
}

\end{document}